# On the mechanism of negative compressibility in layered compounds

E.V. Vakarin [a], A.V. Talyzin [b]

[a] UMR 7575 LECIME ENSCP-UPMC-CNRS, 11 rue P. et M. Curie, 75231 Cedex 05, Paris, France

[b] Department of Physics, Umeå University, SE-901 87 Umeå, Sweden

A mechanism of negative compressibility occurring in compressed layered compounds in the presence of a pressure transmitting fluid medium is discussed within a simple model. It takes into account the excluded volume effects and soft fluid-matrix repulsion. It is demonstrated that a non-monotonic behavior of the interlayer spacing with the applied pressure results from a competition between the applied pressure and the internal one induced by the adsorbed fluid. Recent experimental data on the graphite oxide "structure breathing" under compression in the presence of water are analyzed in the light of these theoretical results.

**Introduction**

"Negative" volumetric compressibility has been previously observed for nanoporous materials, e.g. zeolites or pyrochlores immersed in liquid media [1-5]. The unit cell volume of these materials shows jump-like increase at a certain pressure due to pressure induced insertion of water into pores. Recently we showed that graphite oxide immersed in liquid water medium shows anomalous pressure behavior exhibiting "negative compressibility" and maximum on pressure dependence of cell volume[6].

The unit-cell volume of graphite oxide (GO) pressurized in the presence of water excess continuously increases reaching a maximum at 1.3-1.5 GPa. A sharp downturn in the pressure dependence of the unit-cell volume of the graphite oxide above 1.3-1.5 GPa correlates with solidification of liquid water. The effect is reversible, resulting in a unique breathing of the structure upon pressure variation[6]. The anomaly is explained by pressure induced insertion of water into interlayer space of graphite oxide upon compression at pressures below the unit cell volume maximum and at least partial release of interlayer water back to the inter-grain bulk above the maximum.

Hydrated graphite oxide can be considered as a model material for studies of confinement effects. In a simplest model the hydrated graphite oxide could be considered as system with pack of planar layers and liquid water confined in nano-space between them. The real GO/water system is much more complex. However even this simple model may contribute to understanding of high pressure anomaly observed in GO/water system.

Based on the recent theoretical results[7,8] in this study we propose a qualitative analytical approach to explaining the non-monotonic behavior of the inter-layer spacing in the presence of a pressure transmitting fluid medium.

**Theory**

Model: Graphite Oxide matrix is composed of randomly oriented slit-like pores. If the pores are assumed to be independent, then we can focus on a single pore of variable width $h$. The later varies in response to applied hydrostatic pressure $D$. Immersion the matrix into a liquid (water) induces an internal pressure $P$ on the pore walls (see the sketch below, Figure 1). Here we consider only the normal internal pressure component neglecting the tangential counterpart (which however could be related to possible lateral deformations).

"Dry" matrix: Without considering a microscopic model for the graphite matrix we may assume a simple relaxation process (under the applied pressure $D$) governed by

$$\frac{dh}{dD} = -\alpha D \tag{1}$$

where $\alpha$ is relaxation constant characterizing the matrix rigidity. The magnitude of this constant can be found from a comparison of our theoretical prediction (2) and the experimental data for the dry matrix case. For simplicity we neglect possible

deformations along the pore walls. Solving (1) with appropriate boundary conditions we obtain

$$h(D) = h_f + (h_0 - h_f)\exp(-\alpha D) \qquad (2)$$

where $h_0$ is the initial ($D=0$) pore width and $h_f$ is the final ($D \to \infty$) pore width. As expected, $h(D)$ is a decreasing function of the applied pressure $D$.

Matrix immersed in a liquid: In that case the liquid enters the pore under the external pressure $D$. Depending on the nature of the liquid and its interaction with the pore this process induces the internal pressure $P$ on the pore walls. Therefore the width variation is governed by a difference $D - P$. In analogy with the dry matrix case instead of (2) we may accept

$$h(D, P) = h_f + (h_0 - h_f)\exp(-\alpha[D - P]) \qquad (3)$$

Nevertheless, $P$ depends on $D$ and we have to find a relation $P(D)$. This requires a model for the insertion process. Note that the rigidity constant $\alpha$ could also depend on the presence of the adsorbed liquid, but in the absence of any reliable information on such a dependence we neglect it.

Insertion model: The bulk liquid is modeled as a hard sphere system with the density $\rho_b$ and the hard sphere diameter $\sigma$. In this way we take into account only the packing effects which are dominant at high pressures. The effects of directional attractive forces leading to the formation of networks (specific for hydrogen bonded liquids, such as water) are neglected for the moment. If necessary these phenomena can be taken into account using the methods developed in[9]. In the simplest excluded volume approximation the bulk equation of state is

$$D(\rho_b) = \frac{\rho_b}{1 - b\rho_b} \tag{4}$$

where $b = 2\pi\sigma$. For the liquid-pore interaction we consider a soft repulsion[7,8] neglecting the attractive forces. The latter are responsible for the formation of adsorbed layers at relatively low pressures, while at high pressures the repulsive part of the liquid pore interaction dominates. Therefore, the pore liquid equation of state is[7,8]

$$P(\rho_p) = \frac{\rho_p}{1 - b\rho_p} + \frac{A}{h(D,P)}\rho_p \tag{5}$$

where $A$ is the repulsion magnitude and $\rho_p$ is the pore density. In order to couple these two equations we have to find a relation for the two densities $\rho_p = f(\rho_b)$. As a simplest approximation we accept a linear relation $\rho_b = g\rho_p$. Then from (4) and (5) we obtain a set of equation involving $P(\rho_p = \rho)$ and $D(\rho_b = g\rho)$. This allows us to eliminate $\rho$ obtaining $P(D)$ in the following form

$$P(D) = \frac{D}{g(1+bD) - bD} + \frac{AD}{g(1+bD)}\frac{1}{h(P,D)} \tag{6}$$

Here the first term is essentially of the bulk origin, the second term reflects the pore-bulk coupling. Note that at this level the width $h(P,D)$ is still unknown.

**Results**

Inserting (6) into (3) we obtain a closed equation for $h(D) = h(P(D),D)$. Solving this equation leads to the following dependence of the pore width on the applied pressure (Figure 2). The non-monotonic behavior results from an interplay of internal and external pressures. In the low pressure region the internal pressure is higher and the matrix expands.

In the experimental studies on graphite oxide immersed in water a non-monotonic behavior is also observed, but the maximum at $D = D^*$ is much sharper. Moreover, $D = D^*$ is associated with the freezing of water in the bulk. In this situation we may suppose that the exchange of water between the bulk and the pore stops at $D = D^*$. Then several scenarios are possible. Increasing the internal pressure beyond $D = D^*$ should squeeze out the pore liquid (at least partially). In that case the internal pressure could change. From the experimental point of view we do not know what happens at this point. Of course, our theoretical model is too simple and cannot describe the freezing transition. On the other hand, this is not necessary for our purposes. The only essential point to be taken into account is that upon its freezing the bulk water cannot enter the pore with increasing applied pressure $D$. Therefore, we may accept (as a hypothesis) that $P(D)$ is given by (6) for $D \leq D^*$ and remains $P(D = D^*)$ for $D > D^*$. This piece-vise description leads to the following dependence of the pore width $h$ on the applied pressure $D$ (see Figure 3). The pore expands in the low-pressure region because the internal fluid pressure $P$ is higher than the external one. After reaching a maximum at $D = D^*$ the internal pressure stops to grow (presumably, because the fluid exchange is blocked) and the pore width returns to the usual behavior. Interestingly, that our simple model, that does not take into account the microscopic details of water and the GO matrix, is in qualitative agreement with the experimental results[6] on the water-GO system.

**Conclusion**

In this paper a mechanism of "negative compressibility" occurring in compressed layered compounds in the presence of a pressure transmitting fluid medium is discussed within a simple model. It takes into account the excluded volume effects and soft fluid-matrix repulsion. It is demonstrated that a non-monotonic behavior of the interlayer spacing with the applied pressure results from a competition between the applied pressure and the internal one induced by the adsorbed fluid. In order to reproduce this effect we have to assume that the adsorbed fluid is denser than the bulk one. This is in agreement with an intuitive view that under the external compression some effective "density" (fluid plus matrix) should increase. If the matrix expands in the low-pressure region, then the fluid density in the matrix should increase. Recent experimental data[6] on the graphite oxide "structure breathing" under compression in the presence of water are analyzed in the light of these theoretical results. The theoretical estimations are shown to be in a qualitative agreement with experimental results.

In the recent studies[7,8] it has been demonstrated that the matrix dilatation can induce negative compressibility states for the confined fluid. In this context it would be quite interesting to examine if in the case of GO-water system the confined water also exhibits such a behavior. This issue is left for a future study.

**Figures**

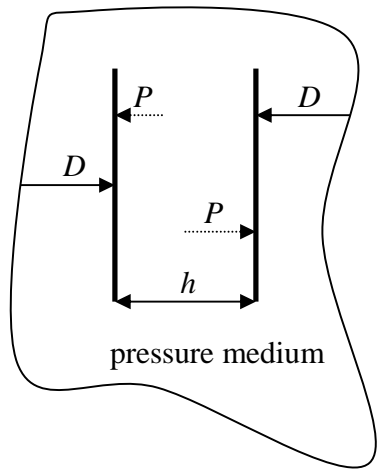

Figure 1. A sketch of the model system. A pore of a variable width $h$ is immersed in a pressure medium. The external pressure $D$ induces the internal pressure $P$ on the pore walls.

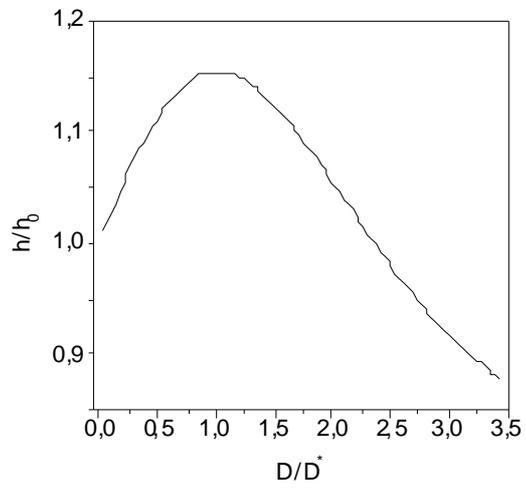

Figure 2 : Relative pore width $h/h_0$ as a function of the relative applied pressure $D/D^*$

The dimensionless parameters are: $h_0/h_f = 11/9$, $b = 1$, $g = 1.1$, $A = 0.5$, $\alpha = 7$.

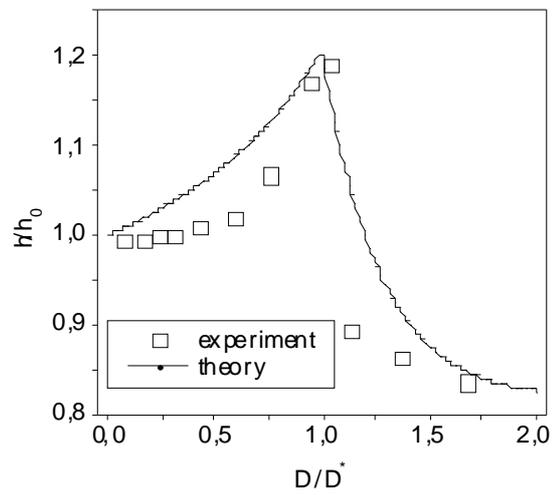

Figure 3 : Relative pore width $h/h_0$ as a function of the relative applied pressure $D/D^*$

The dimensionless parameters are: $h_0/h_f = 11/9$, $b = 1$, $g = 0.94$, $A = 0.1$,

$\alpha = 2.6$.